\documentclass[12pt]{article} 
\usepackage{graphicx}
\usepackage{umlaut}
\usepackage[english]{babel}

\begin{document}
\title{Conductive nanodots on the surface of irradiated CaF$_2$}

\author{Tino Roll, Marion Meier, Henning Lebius and Marika Schleberger \footnote{electronic address: marika.schleberger@uni-due.de}} 

\begin{abstract}
CaF$_2$(111) single crystal surfaces have been irradiated with swift heavy ions under oblique angles resulting in 
chains of nanosized hillocks. In order to characterize these nanodots with respect to their conductivity we
have applied non-contact atomic force microscopy using a magnetic tip. Measurements in UHV as well as under ambient
conditions reveal a clearly enhanced electromagnetic interaction between the magnetic tip and the nanodots.
The dissipated energy per cycle is comparable to the value 
found for metals, indicating that the interaction of the ion with the target material leads to the creation of metallic Ca nanodots on the surface.
\end{abstract}

\maketitle


The irradiation of insulating materials with fast heavy ions can be used to locally modify material properties such as structure \cite{Choudarya08}, resilience \cite{Thibaudau91} or conductivity \cite{Weidinger04}. Probing these modifications with spatially resolving methods is not an easy task since the ion excites the electronic system along its trajectory and heat diffusion limits the extent of the affected volume to a few yoktoliters. At perpendicular incidence the ion creates a hidden track inside the volume of the crystal and a visible modification such as a nanosized crater \cite{Papaleo99} or hillock \cite{Mueller00,Khalfaoui06,Saleh07} occurs only at the entry point. By tilting the beam with respect to the sample surface a much larger part of this hidden track becomes exposed making it easier to detect and characterize the modifications accompanying the track \cite{Akcoeltekin07,Akcoeltekin08}. 

In order to study the nature of the modifications we apply the so-called eddy current microscopy (ECM). This off-spring of atomic force microscopy (AFM) is performed using an oscillating magnetic tip to detect the long-range electromagnetic interactions between tip and surface. The varying magnetization induces an eddy current in a conducting sample giving rise to Joule heat dissipation which is detected as a dissipation of oscillation energy. The contrast mechanism is not yet fully understood, but the method has been shown to be sensitive to the conductivity of a sample \cite{Hoffmann98,Lantz01,Roll08}.

We have chosen CaF$_2$ as a sample because this material is known to form metal Ca colloids upon electron irradiation \cite{Bennewitz94,Reichling96}. A fast heavy ion looses its energy almost exclusively via excitations of the target electrons and not due to collisions with the target atoms. Therefore, anion voids equivalent to metal Ca colloids could be formed along the ion's track due to the strong electronic excitations in the wake of the projectile \cite{Khalfaoui05}.

\begin{figure}[htbp]
\includegraphics[width=8.5cm]{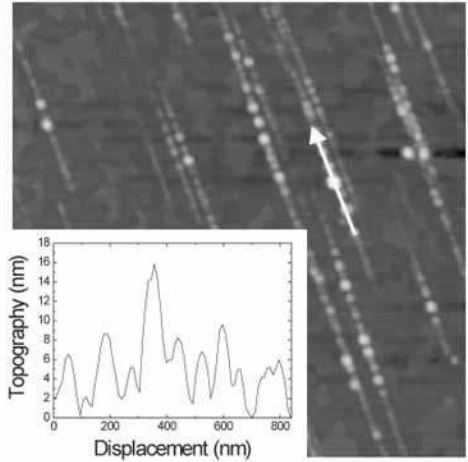}
\caption{Topography image of chains of nanodots on CaF$_2$(111) created by irridation with 93 MeV Pb$^{28+}$ ions. The inset shows a linescan along the arrow, which also depicts the direction of the incoming beam. The height of the nanodots is in the range from 6 to 15~nm. Frame size: $2.5\times2.5~\mu$m$^2$, tip: MESP-HM.}
\label{figure1}
\end{figure}

The CaF$_2$(111) samples (Korth Kristalle, Berlin) were cleaved in air and irradiated with 93 MeV Pb$^{28+}$ ions at the beamline IRRSUD at the GANIL, France. The angle of incidence with respect to the surface was 2$^{\rm{o}}$ and the fluence was typically $\leq1\times 10^{9}$~ions/cm$^2$ to avoid overlapping tracks. 
As can be seen from Figure~(\ref{figure1}), under these conditions every ion produces a chain of separate nanodots aligned along the track of the projectile. The chains are on average 1000 nm in length and the height of the individual dots varies between 6 and 15 nm.

The subsequent AFM measurements were at first performed under ambient conditions (Dimension 3100/NanoScope V). The AFM was operated in the amplitude modulation (AM) mode \cite{Lee06}. In this mode the electromagnetic interaction of the tip with the sample is detected via a phase shift between the excitation amplitude and the oscillating cantilever. In order to separate these long-range forces from the topography, the AFM was operated in the LiftMode$^{\sf TM}$ at various distances. As ECM probes, cantilevers with a magnetically coated tip (VEECO MESP-HM, Cr/Co, $f_0=64$~kHz, $k=3$~N/m) were used. The tip radius is less then $r_{tip}=10$~nm, and the magnetic moment is $m\geq3\times10^{-13}$~Am$^2$. The cantilever oscillates with an amplitude of $A=15$~nm and the conversion of the measured phase shift $\varphi$ into an energy loss per oscillation cycle can be calculated as follows \cite{Cleveland98}
\begin{equation}
	\overline{E_{ts}}= \frac{1}{2}\frac{kA^2\omega_0}{Q}\left[\left(\frac{A_0}{A}\right)sin\varphi-1\right]
\end{equation}
with A$_0 \approx$ A in LiftMode$^{\sf TM}$.  

\begin{figure}[htbp]
\includegraphics[width=8.5cm]{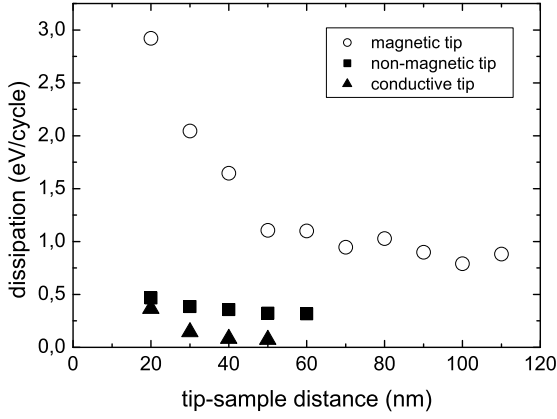}
\caption{Distance dependence of the phase signal for cantilever with a magnetic (squares), a non-magnetic (circles) and a conductive tip (triangles).}
\label{figure2}
\end{figure}

Figure~(\ref{figure2}) shows the dissipated energy per oscillation cycle determined from the $rms$-roughness of the phase images, as a function of the tip sample distance. The open circles represent the signal acquired with a cantilever with a magnetic tip, whereas the triangles represent the data acquired using a cantilever with a conductive tip (Nanosensors, NCHPt, $f_0=280$~kHz, $k=42$~N/m). The latter signal vanishes at a distance of 60~nm, while the signal originating from the interaction with a magnetic tip is still about 1.1~eV/cycle a this distance. In the range from 20~nm up to 50~nm the signal for the cantilever with a magnetic tip decreases linearly from 2.9~eV/cycle down to 1.1~eV/cycle. At distances larger than 50~nm the signal decreases only slightly by $\approx$~0.2~eV/cycle. For comparison we have added the data measured with a conventional Si tip (Nanosensors, QNCHR, $f_0=300$~kHz, $k=42$~N/m, squares) which also gives a rather weak phase signal. The large signal obtained exclusively with the magnetic tip at small distances as well as the non-vanishing signal at large distances indicates a significant electromagnetic interaction which is present at the locations of the hillocks only. 

The phase signal in AFM is known to be sensitive to various physical properties of the surface and it is rather difficult to determine the energy dissipation in a quantitative and unambiguous way. We therefore introduced the sample into a UHV setup (RHK AFM/STM UHV 7500, $p_{base}\leq3\times10^{-10}$~mbar) and used another detection scheme known as frequency modulation detection (FM-AFM)\cite{Albrecht91}. In this mode one feedback loop controls the separation between the tip and the sample by keeping the frequency shift $df$ at a certain value. This signal contains the topographical information. A second feedback control loop keeps the oscillation amplitude constant by replacing any dissipated energy. Measuring this quantity we can determine the average energy dissipation due to non-conservative (dissipative) tip-sample interaction directly and independently of the topography in the same scan. Of course,  the intrinsic dissipation of the lever due to internal friction has to be accounted for. We calculate this intrinsic dissipated energy of the freely oscillating cantilver according to 
\cite{Anczykowski99} 
\begin{equation}
	E_{0}= \frac{\pi kA^{2}}{Q}.
\end{equation}
and find an energy dissipation of typically $E_0\approx8$~eV per oscillation cycle ($A=$~34~nm, $Q\approx$~10000).

\begin{figure}[htbp]
\includegraphics[width=8.5cm]{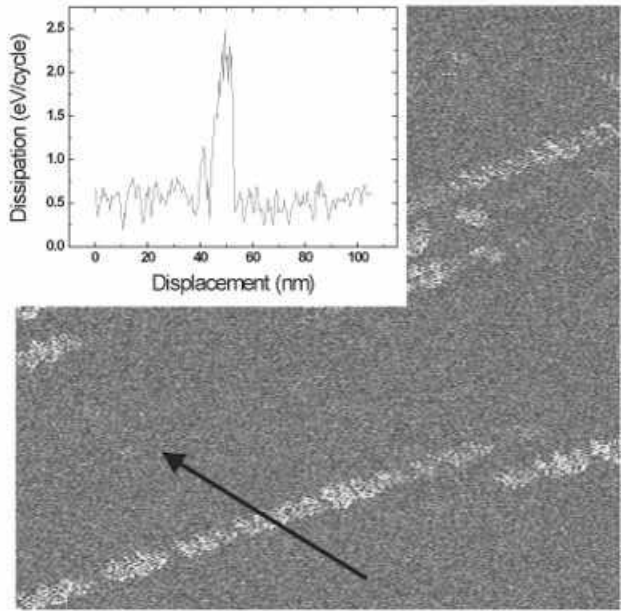}
\caption{(a) Dissipation image of the CaF$_2$ sample. Image has been acquired using $\gamma = -1.6$~fNm$^{1/2}$. The frame size is $215 \times215~$nm$^2$. Dissipation due to tip-sample interaction is about 0.5~eV/cycle when scanning over the non-irridiated areas. Oscillating the tip above the hillocks lead to a dissipation of 2~eV/cycle. The double-peak structure is due to a multiple tip.}
\label{figure3}
\end{figure}

To reduce the influence of possible residues on the surface the irridiated CaF$_2$ sample was heated in a load lock chamber at 400 K for 4 h and subsequently transferred into the AFM. In order to compensate for any long ranged electrostatic interactions between tip and sample a bias voltage of $U_{Bias}=1.8$~V was applied during imaging. The topography images revealed basically the same morphology as shown in Figure~(\ref{figure1}), the dissipation signal recorded at the same time is shown in Figure~(\ref{figure3}). The image was taken at a frequency shift of $df=-6$~Hz, which is equal to a normalized frequency shift of $\gamma = \frac{\Delta f}{f_0} k A^{3/2} =-1.6$ fNm$^{1/2}$, using a cantilever with a magnetic tip at room temperature. The nanodots can be clearly identified in the dissipation signal as the signal is significantly enhanced at the location of the dots. Because pronounced topographic features are known to give rise to artefacts in dissipation images, both scanning directions and the error signal of the two control loops, oscillation amplitude and frequency shift, were checked carefully and no indications for such artefacts were detected. 
Note, that using an uncoated, i.e. non-magnetic tip and a cantilever with a conductive tip, the dissipation image exhibited no contrast at all. These experiments were done at the same normalized frequency shift to ensure comparability. Thus, the dissipative interaction must again be due to the magnetic nature of the oscillating tip, presumably giving rise to an eddy current in the conductive hillocks. 

Using the voltages V$_{exc}$ and V$_{exc,0}$ close and far away from the surface, respectively one can determine the dissipated energy per oscilliation cycle using the following formula \cite{Anczykowski99}:
\begin{equation}
	E_{ts}=E_{0}\left(\frac{V_{exc}}{V_{exc,0}}-\frac{f}{f_0}\right)
\end{equation}
Taking the data from our measurements we find about 0.5~eV/cycle at the (non-irradiated) substrate and 2.0~eV/cycle at the position of the hillocks. These values are slightly lower than the values found under ambient conditions, probably due to the reduced residues and compensation of electrostatic interaction. The overall dissipated power is in the pJ range and in good agreement with data acquired by using a tip with a magnetic moment of about $10^{-9}$~Am$^2$ on semiconducting samples \cite{Lantz01}.

Not much is known yet about the inner structure of the nanodots produced in ion irradiation experiments. The modified volume is too small to apply spectroscopic techniques and in addition the material is very susceptible to any kind of irradiation. In general, it is assumed that for the hillocks creation melting of the material is a prerequisite \cite{Toulemonde99,Akcoeltekin08}. On the other hand, 
non-amorphizable ionic materials are known to show a significant volume 
swelling under irradiation. In CaF$_2$ samples irradiated with 
C$_{60}$ clusters (resulting in a higher energy loss), calcium 
inclusions have been found \cite{Jensen98}. In our case, i.e. at 
glancing angles, it may be the surface playing a major role for the 
formation of nanosized hillocks devoid of fluorine. The creation of metal Ca nanodots would be consistent with the ECM data presented here. The conductivity of pure Ca is however rather low and the dissipation signal is relatively high, as compared with data obtained from other samples. Thus, the details of the contrast mechanism remain yet to be elucidated.

In conclusion, we have demonstrated that the irradiation of CaF$_2$ surfaces under oblique angles produces long tracks consisting of a series of nanosized regions which are different from the surrounding non-irradiated CaF$_2$ matrix. These nanodots do not only show up as hillocks in conventional topographic AFM images but give rise to a substantial dissipation of energy when probing the electromagnetic interaction. This is a clear indication that the nanodots are indeed conductive.


\section*{Acknowledgement}

Irradiations were performed at the beamline IRRSUD at GANIL, Caen, France.
We thank I. Monnet for her support and fruitfull discussions.
Financial support from the Deutsche Forschungsgemeinschaft through SFB 616 "Energy dissipation at surfaces" is gratefully acknowledged.

\section*{References}

\newpage

\end{document}